\documentclass[prd,aps]{revtex4}
\usepackage{amsmath,amssymb,amsfonts,dcolumn,color,graphicx,graphics,latexsym,placeins,epsfig}
\usepackage[compatibility=false]{caption}
\usepackage{subcaption}
\usepackage{natbib}
\newcommand{\be}{\begin{equation}}
\newcommand{\ee}{\end{equation}}
\newcommand{\ba}{\begin{eqnarray}}
\newcommand{\ea}{\end{eqnarray}}

\newcommand{\lb}{\left(}
\newcommand{\rb}{\right)}

\begin{document}

%\preprint{LIGO-P1800035}

\title{\Large \bf Gravitational waves from quasinormal modes of a class of Lorentzian wormholes}
\author{S. Aneesh${}^1$, Sukanta Bose${}^{2,3}$ and
Sayan Kar ${}^{4}$}
\email{aneesh.s306@gmail.com,sukanta@iucaa.in,sayan@phy.iitkgp.ernet.in}
\affiliation{${}^1$Department of Physics, Indian Institute of Technology, Kharagpur, 721 302, India}
\affiliation{${}^{2}$ Inter-University Centre for Astronomy and Astrophysics,
Post Bag 4, Ganeshkhind, Pune 411007, India}
\affiliation{${}^{3}$Department of Physics \& Astronomy, Washington State University, 1245 Webster, Pullman, WA 99164-2814, U.S.A.}
\affiliation{${}^{4}$ Department of Physics {\it and} Center for Theoretical Studies \\Indian Institute of Technology, Kharagpur, 721 302, India}

\begin{abstract}
Quasinormal modes of a two-parameter family of Lorentzian wormhole spacetimes,
which arise as solutions in a specific scalar-tensor theory 
associated with braneworld gravity, are 
obtained using standard numerical methods. Being solutions in a
scalar-tensor theory, these wormholes can exist with matter
satisfying the Weak Energy Condition. %Assuming that the end-state ringdown of some astrophysical merger process to be such a wormhole,
If one posits that the end-state of stellar-mass binary black hole mergers, of the type observed in GW150914, can be these wormholes, then we show how their properties can be measured from their distinct signatures in the gravitational waves emitted by them as they settle down in the post-merger phase from an initially perturbed state. We propose that their scalar quasinormal modes correspond to the so-called breathing modes, which normally arise in gravitational wave solutions in scalar-tensor theories. We show how the frequency and damping time of these modes depend on the wormhole parameters, including its mass. We derive the mode solutions and use them to determine how one can measure those parameters when these wormholes are the endstate of binary black hole mergers. Specifically, 
%by using the parameters of GW150914, we estimate the statistical errors of one of the characteristic wormhole parameters. We 
we find that if a {\em breathing} mode is observed in LIGO-like
detectors with design sensitivity, and has a maximum amplitude equal to 
that of the {\em tensor} mode that was observed 
of GW150914, then for a range of values of the wormhole parameters, we will be able to discern it from a black hole. If in future observations
we are able to confirm the existence of such wormholes, we would, at 
one go, have some indirect evidence of a modified theory of gravity 
as well as extra spatial dimensions.  

\end{abstract}

%{Classical general relativity, Exact solutions}

\maketitle

\section{\bf Introduction}

\noindent Lorentzian wormholes have been around as theoretical 
constructs ever since the idea of the Einstein-Rosen (ER) bridge
was born in 1935 \cite{einstein}. Among many of Einstein's ideas 
and predictions, 
gravitational waves (GW) and the cosmological constant are 
a part of reality today \cite{gwobs,gwobs1,lambda}, 
but the Einstein-Rosen bridge 
and its progeny -- the wormholes-- are yet to see the light of day
in the real universe.       

\noindent Subsequent to the ER article and about a couple of decades later,
Misner and Wheeler, in their paper on classical physics
as geometry \cite{mw}, first introduced the term {\em wormhole}. Later,
through the work of Ellis  \cite{ellis}, Bronnikov \cite{bronnikov}, 
Morris, Thorne and Yurtsever \cite{mty}, Morris and Thorne \cite{mt}, Novikov \cite{novikov}, Novikov and Frolov \cite{novikov1}, Visser \cite{visser}
and many others \cite{wothers1,wothers2,wothers3,wothers4,wothers5,wothers6,wothers7,wothers8,wothers9,wothers10, wothers11}, the wormhole idea was further 
developed with numerous examples as well as 
enquiries into the intriguing possibilities that may arise
with wormholes (eg. time machines!~\cite{mty, novikov, novikov1}). Even today, the term {\em wormhole}, 
does appear almost every day,
in one article or the other, in the daily list of submitted articles 
in preprint archives.

\noindent Wormholes are, in some sense, {\em good spacetimes}!
They do not have {\em horizons} or {\em singularities}, which 
make things interesting as well as difficult. But the absence of
horizons or singularities for wormholes comes at a heavy cost. 
The matter required to have a wormhole violates the so-called
{\em energy conditions} \cite{hawkellis,wald}, atleast in the context of
General Relativity. Wormholes seem to require exotic matter -- i.e.,
matter for which energy density can become negative in some frame of
reference.

\noindent Is there a way to avoid this impasse? Many resolutions have been suggested
in the past~\cite{wecviolation1,wecviolation2,wecviolation3, hochberg, wecviolation4,wecviolation5,roman,
wecviolation6,wecviolation7,wecviolation8,wecviolation9,wecviolation10, canfora1,canfora2} Among them, one avenue is to look into
modified theories of gravity where additional degrees of freedom
(e.g., say a scalar field) have a role to play. In the old Brans-Dicke
idea \cite{bransdicke,bransdicke1}, the scalar field replaced the gravitational constant $G$. 
In later versions and the most recent ones, the scalar field can 
actually arise via the presence of extra spatial dimensions \cite{kannosoda}. 
A well-known model that exploits this is the on-brane gravity \cite{kannosoda} arising in the 
so-called two-brane model of Randall--Sundrum \cite{randall},
wherein the scalar field is related to the inter-brane-distance.
Thus, we and our wormhole would be on one such 3-brane and a
scalar field, which is not quite `matter', would provide the
required negativity so that the `convergence condition' is
violated (as it must be for wormholes) \cite{hawkellis}, 
but the `matter' threading the 
wormhole is usual matter, with all the desired properties. The above
line of thought was exploited to construct a class of wormholes with 
matter satisfying the
energy conditions, in work done recently by one of the authors here
(along with others)\cite{sksl,sksl1}. In a way, therefore, the existence of the
wormhole would therefore provide support to an alternative theory
of gravity, as well as to the existence of extra dimensions!

\noindent How then does one show that such a wormhole does indeed exist?
Motivated by recent detections of gravitational waves
at LIGO and Virgo~\cite{gwobs,gwobs1}, we explore whether there is any meaning to
a proposal that the final state of some violent collision of
neutron stars and/or black holes might result in a wormhole 
of the type we mention above or, more, realistically, its rotating
version. We do not have any model
which shows that a wormhole may indeed emerge in such a collision.
However, such a suggestion is not entirely new. (See~\cite{ringdown1,damour, taylor,ringdown2,ringdown3,echoes,rezzolla}
for  earlier work as well a more recent one
on GW signals from wormholes.)
All we can say, is that, by studying the ringdown and the 
quasinormal modes (which we find here), we can, through a
comparison with observational data, estimate the error bounds
in the parameters which define the wormhole and appear in the
quasi-normal modes. The values of the wormhole parameters
may be constrained by other means such as lensing
or time-delay. Thereafter, we can say, to what
extent, through gravitational wave observations we can
constrain the merged object to be a wormhole. 
It is true, however, that the BBH (binary black hole) GW signals observed so far are all
consistent with the merger of two Kerr black holes to another Kerr
black hole, but the extent to which mergers of objects that are not
Kerr black holes could resemble these signals is yet to be established [2,3].
%Though our proposal may sound outrageous, there is some element of
%truth in it, which is further strengthened by the fact that, despite
%all efforts, one is not fully sure, about the precise nature of the
%merged object, in almost all instances of GW observations which have appeared till date.

\noindent Our paper is organised as follows. In the next section,
we briefly recall the spacetime and the theory for which this is
a solution. Thereafter, in Section III,  we set up the search for massless
scalar  quasinormal modes, in this background geometry. We 
try to justify how
these scalar QNMs could precisely be those for the so-called 
{\em breathing mode}. We solve for QNMs numerically, find them
and demonstrate their characteristics through various plots and
analysis. In Section IV, we demonstrate how one can estimate the
errors in the wormhole parameters (more precisely, one parameter)
using inputs from GW observations. Finally, in Section V, we
sum up and provide possible avenues for future work. In the rest of the paper we will use units in which $G=1$ and $c=1$.

\noindent 

\section{The class of wormhole spacetimes}

\noindent Let us begin with the modified theory of gravity, in which, our 
wormhole
is a solution. As stated in the Introduction, this is a scalar-tensor theory
of a specific type. It arises as a theory on the four dimensional 
3-brane timelike  hypersurface in a five-dimensional background.
We have two 3-branes separated by a distance in extra-dimensional space--
the inter-brane distance is associated with the scalar, in our low-energy,
effective, on-brane scalar tensor theory of gravity. The subsections
below briefly recall the theory as well as the wormhole solution.

\subsection{Scalar tensor gravity, field equations, wormhole solutions} 

\noindent The field equations for the on-brane scalar-tensor theory of gravity 
are given by \cite{kannosoda},
\begin{align}
    R_{\mu\nu}=\frac{\bar\kappa}{l\Phi}\left( T^{b}_{\mu\nu}-\frac{1}{2}g_{\mu\nu}T^{b}\right) +\frac{\Omega(\Phi)}{\Phi^{2}}\nabla_{\mu}\Phi\nabla_{\nu}\Phi+\frac{1}{\Phi}\lb\nabla_\mu\nabla_{\nu}\Phi+\frac{1}{2}g_{\mu\nu}\Box\Phi\rb
\end{align}
where $T^b_{\mu\nu}$ is the stress energy tensor on the 3-brane
(labeled as the ``$b$'' brane in \cite{kannosoda}) and $\Phi$ is the scalar field which satisfies the field equation,
\begin{align}
    \Box\Phi=\frac{\bar\kappa}{l}\frac{T^{b}}{2\Omega+3}-\frac{1}{2\Omega+3}\frac{d\Omega}{d\Phi}\nabla^{\alpha}\Phi\nabla_{\alpha}\Phi
\end{align}
$l$ is the bulk curvature radius and $\bar \kappa$ is related to the higher
dimensional Newton constant.
The coupling function $\Omega (\Phi)$ can be expressed in terms of the scalar 
field as,
\begin{align}
    \Omega(\Phi)=-\frac{3\Phi}{2(1+\Phi)}
\end{align}
The scalar field $\Phi$, as mentioned before, is associated with the inter-brane distance in the bulk.
It has to be non-zero, positive and finite in order to have a meaningful two-brane model.
$T^b$ is the trace of the stress-energy on the ``$b$'' 3-brane, embedded in a five dimensional
bulk spacetime. The above field equations are for the scalar-tensor theory on this so-called $b$-brane.  For more details about the theory, the reader is referred to \cite{kannosoda}. 

\noindent In the above-mentioned theory, we now consider a static, spherically symmetric wormhole 
solution of the field equations with a vanishing Ricci scalar. 
Such a solution has been shown to be given by \cite{sksl} (see
earlier work in {\cite{R3,R3'}),
    \begin{align}
        ds^{2}=-\left(\kappa+\lambda\sqrt{1-\frac{2M}{r}}\right)^{2}dt^{2}+\frac{dr^{2}}{1-\frac{2M}{r}}+r^{2}\left(d\theta^{2}+\sin^{2}\theta 
        d\phi^{2}\right)
    \end{align}
where $\kappa$, $\lambda$ are non-zero, positive constants. Note that our wormhole has two parameters: $M$,
a measure of the throat radius ($2M$ here, like in Schwarzschild) and $\frac{\kappa}{\lambda} >0$ which, being non-zero, signals the absence of a horizon. 

\noindent The Jordan frame scalar field $\Phi$ takes the form,
    \begin{align}
        \Phi=\left(\frac{c_1}{M\lambda}\ln\frac{2 r'q+M}{2r'+M}+c_2\right)^2-1
    \end{align}
where $r=r'(1+\frac{m}{2r'})^2$ ($r'$ is the isotropic coordinate), $q=\frac{\kappa+\lambda}{\kappa-\lambda}>1$ and $c_1$, $c_2$ are constants of integration, in this solution. Assuming $c_2=0$ one can show that the WEC can hold under specific choices of the various parameters \cite{sksl}. 
Note that the timelike or null convergence
condition is indeed violated, as it should be, in order to ensure that the
spacetime is a  wormhole. However, the required matter satisfies the 
WEC. For more details about the solution, the stress-energy of the matter that supports the
wormhole, as well as the WEC see \cite{sksl}. 

\noindent If $q<0$ (i.e. $\frac{\kappa}{\lambda} <1$), the scalar field solution remains similar in its functional form. In order to have a finite, non-zero radion scalar and also ensure that the WEC holds,
we need $c_2\neq 0$, as well as additional constraints on the parameters.

\subsection{Scalar field propagation}
\noindent How does a massless scalar field (not necessarily the scalar $\Phi$ in the theory mentioned above, but any generic scalar field) 
propagate in the above-mentioned background spacetime?
Such a scalar field, as we show later, is related to the
perturbations of the Brans-Dicke scalar $\Phi$, which we
introduced in the previous subsection. In addition, there
are also gravitational perturbations which we do not fully consider here.
Our problem therefore reduces to 
solving a massless Klein-Gordon equation in a fixed, curved background, i.e.
\begin{align}
    \Box\Psi=0
  \end{align}
Since the background spacetime is 
spherically symmetric and static, we can decompose $\Psi$ in terms of the spherical 
harmonics,
\begin{align}
    \Psi(t,r,\theta,\phi)=\sum_{l=0}^{\infty}\sum_{m=-l}^{l}\frac{\psi_{lm}(r)}{r}e^{-i\omega t}Y_{lm}(\theta,\phi)
\end{align}
By inserting this ansatz in the Klein-Gordon equation we get the equation satisfied by each mode,
\begin{align}
    fh\frac{d^2\psi_{lm}}{dr^2}+\frac{1}{2}(hf'+fh')\frac{d\psi_{lm}}{dr}+\omega^2\psi_{lm}=\frac{rhf'+f(2l(l+1)+rh')}{2r^2}\psi_{lm}
\end{align}
where have defined $f(r)=-g_{tt}$ and $h(r)=(g_{rr})^{-1}$.
We can rewrite the above equation by introducing the tortoise coordinate $r_*$ defined as,
\begin{align}
    \frac{dr_*}{dr}&=(fh)^{-1/2}
    %r_{*}(r)&=\int_{2M}^{r}\sqrt{-\frac{g_{tt}(r')}{g_{rr}(r')}}dr'
\end{align}
The above equation can be integrated to obtain an analytical expression for the tortoise
coordinate, given as, 
\begin{align}\label{rstar}
r_* = \frac{M}{\lambda} \left \{ \frac{2(p-\beta)(2p-\beta)}{(p^2-1)\left [ (p-\beta)^2-1 \right ]} + 4 \frac{\ln \frac{\beta}{p}}{(p^2-1)^2} + \frac{(p-2)\ln (1-p+\beta)}{(p-1)^2} -\frac{(2+p) \ln (1+p-\beta)}{(1+p)^2} \right \}
\end{align}
where $p=\frac{\kappa}{\lambda}$, $\beta= p+\sqrt{1-\frac{2M}{r}}$. The asymptotic regions
of the wormhole correspond to $r_*\rightarrow \pm \infty$. We need to use $\pm r_*$ in order
to cover the two asymptotic regions connected by the throat, which
requires a thin shell joining two copies of the same geometry  \cite{visser}. In some sense, $r_*$ is, therefore, like
the proper radial distance $l$ which also ranges from $-\infty$ to $+\infty$ with $l=0$
being the throat.

\noindent With the above definition and details about $r_*$ one gets at the
following equation for $\psi_{lm}$,
\begin{align}\label{mastereqn}
    \frac{d^2\psi_{lm}}{dr_*^2}+[\omega^2-V_{l}(r)]\psi_{lm}=0       \end{align}
where $V_{l}(r)$, called the effective potential is given by,
\begin{align}
    V_{l}(r)=&\frac{M\lambda}{r^3}\sqrt{1-\frac{2M}{r}}\left(\kappa+\lambda\sqrt{1-\frac{2M}{r}}\right)\nonumber \\&+\frac{1}{r^2}\left(\kappa+\lambda\sqrt{1-\frac{2M}{r}}\right)^2\left(\frac{M}{r}+l(l+1)\right)
\end{align}
This effective potential as a 
function of $r_*$ is symmetric about $r_*=0$ ($r=2M$, the throat) with a double-hump structure and goes to zero 
in the asymptotic regions. The plots below (Figure 1) show the variation of the potential with 
the different parameters that appear in it.
\begin{figure}[h]
 \centering
 \begin{subfigure}[b]{0.5\textwidth}
 	\centering
 	\includegraphics[width=\textwidth]{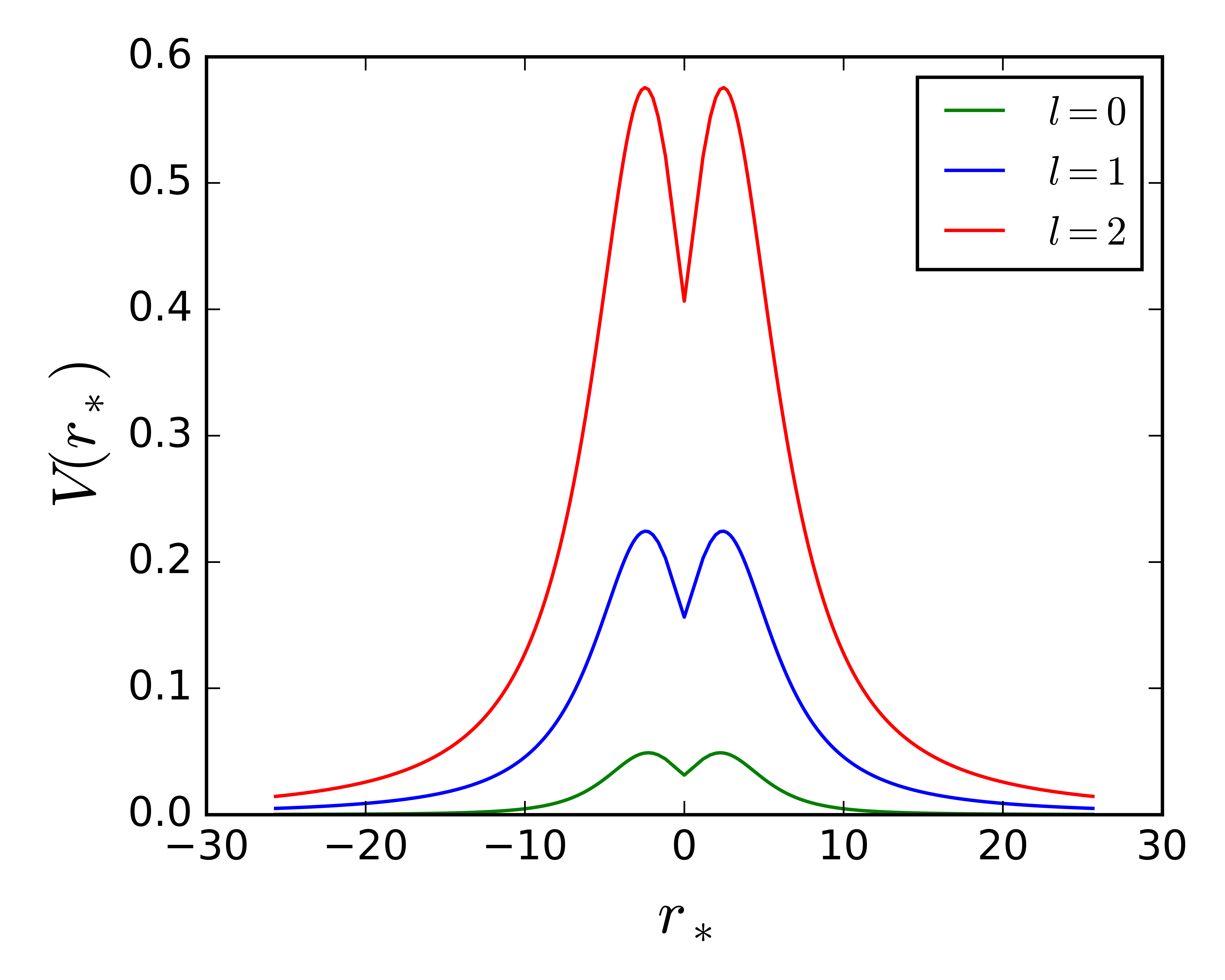}
 	\caption{$\kappa=\lambda=0.5, M=1$}
 	%\label{fig:sub1}
 \end{subfigure}%\hspace{-1em}
 \begin{subfigure}[b]{0.5\textwidth}
 	\centering
 	\includegraphics[width=\textwidth]{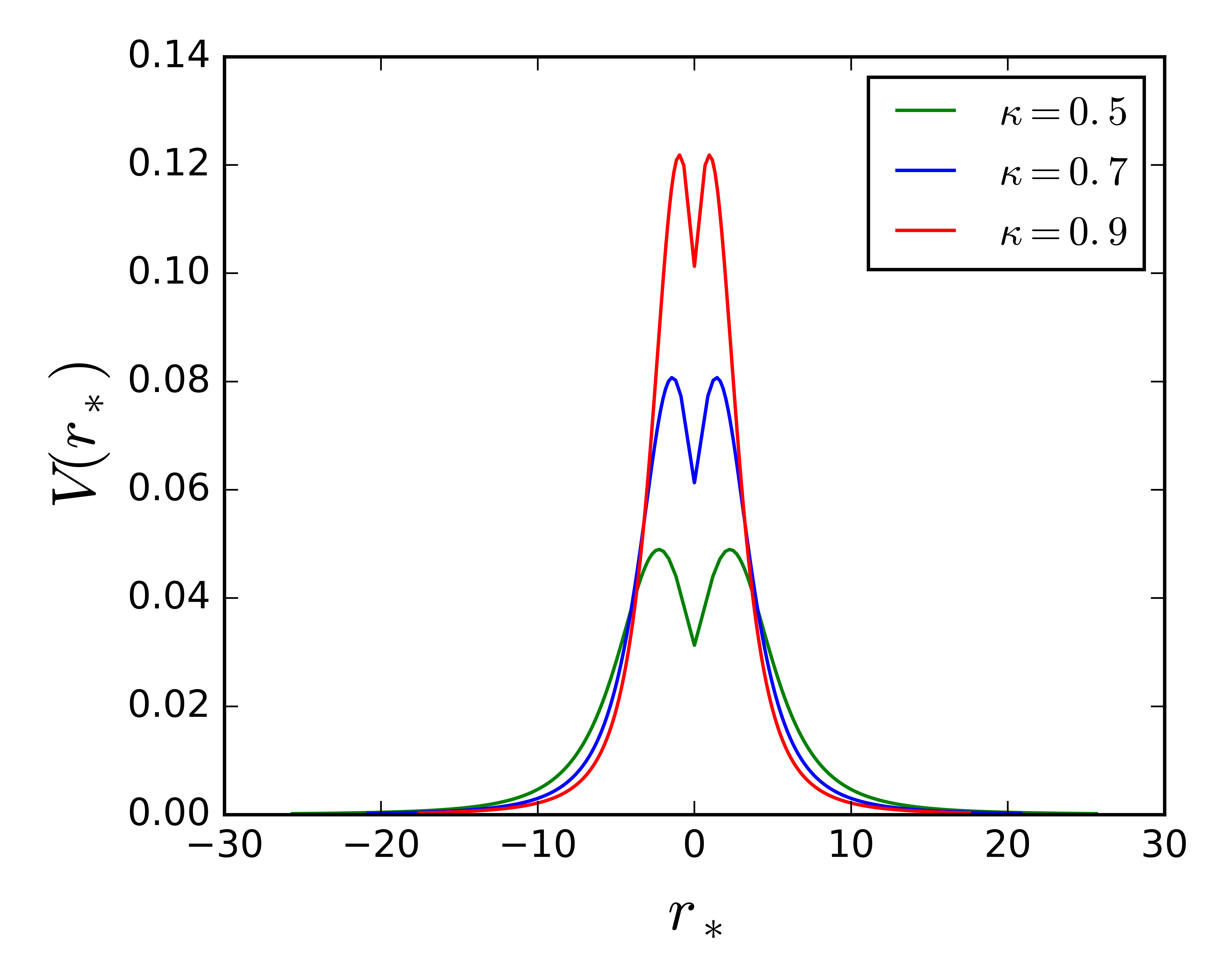}
 	\caption{$l=0, \lambda=0.5, M=1$}
 	%\label{fig:sub2}
 \end{subfigure}
 \caption{Variation of effective potential with the parameters 
$\kappa$, $\lambda$ and $l$. (a) $\kappa=\lambda=0.5, M=1$, (b) 
$l=0$, $\lambda=0.5$, $M=1$.}
 \label{fig:test}
\end{figure}

\subsection{The breathing mode}
In the neighborhood of the detector, the background spacetime is flat
and we consider the perturbation of a flat Minkowski background and the 
constant background scalar field,
\begin{equation}
        g_{\mu\nu}=\eta_{\mu\nu}+h_{\mu\nu},\;\;\;\;    \Phi=\Phi_0(1+\epsilon)
\end{equation}
where $\Phi_0$ is a constant. The field equations become
\begin{eqnarray}
  &\Box (-h_{\mu\nu}+\eta_{\mu\nu}\epsilon)= \frac{2{\bar\kappa}}{l \Phi_0} T_{\mu\nu}\\
&\Box \epsilon=\frac{\bar\kappa}{l{\Phi}_1} T
\end{eqnarray}
where ${\Phi}_1 = \frac{3\Phi_0}{1+\Phi_0}$.
Here the choice of gauge is 
\begin{equation}
  \partial_{\nu}\bar h^{\mu\nu}=\partial^{\mu}\epsilon
\end{equation}
where ${\bar
  h}^{\mu\nu}$ is the trace-reversed metric perturbation. In vacuum, we have $\Box h_{\mu\nu}=0=\Box\epsilon$ and in the transverse traceless gauge, we get
\begin{equation}
h_{\mu\nu}=\begin{bmatrix}
0&0&0&0\\
0& h_+-\epsilon_0&h_\times&0\\
0&h_\times&-h_+-\epsilon_0&0\\
0&0&0&0
\end{bmatrix}e^{i\omega(t-z)}
\end{equation}
for a plane wave propagating in the $z$-direction. The scalar field is $\Phi=\Phi_0(1+\epsilon)$ where
\begin{equation}
  \epsilon=\epsilon_0 e^{i\omega(t-z)}
\end{equation}
Due to the presence of the scalar field, there is an additional polarization in the
gravitational wave, which is known as the breathing mode \cite{breathing}. 
Usually, if we consider a massive scalar, then we also have a longitudinal
mode. However, in our work here, we consider only a massless scalar.

\noindent In a curved background, the equation for scalar field 
perturbation is,
\begin{equation}\label{scalar-eqn}
 \Box(\delta\Phi)=g^{\mu\nu}\delta\Gamma^\alpha_{\mu\nu}\partial_\alpha\Phi+h^{\mu\nu}A_{\mu\nu}-B \,\delta\Phi- C^\alpha \partial_\alpha (\delta\Phi)
\end{equation}
where we have defined,
\begin{eqnarray}
 A_{\mu\nu} &=& \frac{\Omega'}{2\Omega+3}\Phi_{;\mu}\Phi_{;\nu}+\Phi_{;\mu;\nu}\\
 B &=& \frac{2\bar{\kappa}}{l}\frac{\Omega'T^b}{(2\Omega+3)^2}+\frac{\Phi^{;\alpha}\Phi_{;\alpha}}{2\Omega+3}\left(\Omega''-\frac{2(\Omega')^2}{2\Omega+3}\right)\\
 C^\alpha &=& \frac{2\Omega'}{2\Omega+3}\Phi^{;\alpha}
\end{eqnarray}
Here a prime denotes a derivative w.r.t $\Phi$ and we have assumed that there is no fluctuation of $T^b$, the trace of matter stress energy (i.e. $\delta T^b=0$). A similar equation in the Einstein frame
is obtained in \cite{zimmerman}. If we can make an infinitesimal gauge transformation ($h_{\mu\nu}\rightarrow h_{\mu\nu}+\xi_{\mu;\nu}+\xi_{\nu;\mu} \,, \delta \Phi \rightarrow \delta \Phi + \xi^\mu\partial_\mu \Phi$ generated by $x^\mu \rightarrow x^\mu - \xi^\mu$)  which satisfies,
\begin{equation}\label{gauge}
  \left \{ \Box\xi^\alpha-\xi^\mu R^{\alpha}_{\;\;\mu} - B \, \xi^\alpha \right \}\Phi_{;\alpha}+\xi^{\mu;\nu}\left(2 A_{\mu\nu}-\Phi_{;\mu}C_\nu\right)-C^\alpha\Phi_{;\mu;\alpha}\xi^\mu=C^\alpha\partial_\alpha\delta\Phi-h^{\mu\nu}A_{\mu\nu}+B \, \delta\Phi-(h^{\alpha\nu}_{\,\,\,\, ;\nu}-\frac{1}{2}h^{;\alpha})\Phi_{;\alpha}
\end{equation}
where $\xi^\alpha$ is the gauge function, the equation for scalar field perturbations reduces to,
\begin{eqnarray}
\Box(\delta\Phi)=0
\end{eqnarray}
\iffalse
\noindent Since the background scalar field is spherically symmetric $\phi_{;\alpha}=(0,\partial_r\phi,0,0)$. Thus the second term in the gauge condition
is $\xi_\mu R^{r\nu\mu}_{\;\;\;\;\nu}$, which will vanish if we choose $\xi_r=0$. Also, this choice will eliminate the second derivative term in the first term ($\Box\xi^\alpha\phi_{;\alpha}$). Since $A_{\mu\nu}$ is also diagonal, the resulting equation will be of the form,
\begin{eqnarray}  f_1(r)\partial_t\xi^t+f_2(r)\cot\theta\partial_\theta\xi^\theta+f_3(r)\partial_\phi\xi^\phi=f_4(h;\delta\phi;x^\mu)
\end{eqnarray}
Which can be easily solved by choosing $\xi^\theta=0=\xi^\phi$ and,
\begin{equation}
\xi^t=\int dt \frac{f_4(h;\delta\phi;x^\mu)}{f_1(r)}
\end{equation}
\fi
It can be easily shown that  (\ref{gauge}) always admits a solution. For example, since the background spacetime and the scalar field are static and 
spherically symmetric, we may choose the gauge function to be 
$\xi^\mu=(\xi^t,0, 0,0)$. With this choice, it turns out that all
second derivative terms in the equation for $\xi^\mu$ vanish and (\ref{gauge}) reduces to the form $\frac{\partial \xi^t}{\partial t}=f(h_{\mu\nu};\delta\Phi;x^\mu)$, which can always be integrated.

\noindent Since the scalar perturbation obeys the Klein-Gordon equation in the fixed 
background metric, the QNMs calculated may correspond to the breathing mode
mentioned earlier.
%The equation (\ref{gauge}) puts only one constraint 
%on the gauge and we can use the remaining three gauge degrees of 
%freedom to simplify the equation for the tensor perturbations. \par
% In Brans-Dicke theory the coupling function $\omega$ is a constant and thus (43) reduces to the KG equation by a simple guage transformation $(\xi^\mu_{\;\; ;\mu}=-h/2)$\\%
%The equation for the metric perturbation takes the form,
%\begin{eqnarray}
 %\Box h_{\mu\nu}+h_{;\mu;\nu}-h_{\lambda\mu;\nu}^{\;\;\;\;\;\; ;\lambda}-h_{\lambda\nu;\mu}^{\;\;\;\;\;\; ;\lambda}=\bar \kappa h^{\alpha\beta}T_{\alpha\beta}g_{\mu\nu}+h_{\mu\nu}T+\frac{2\delta\phi}{\phi}(T_{\mu\nu}-\frac{1}{2}g_{\mu\nu}T) \nonumber \\-\frac{2\omega}{\phi}(\phi_{,\nu}\phi_{,\mu}(\frac{\omega'}{\omega}-\frac{2}{\phi})+\phi_{,\mu}\delta\phi_{,\nu}+\phi_{,\nu}\delta\phi_{,\mu})+\frac{2\delta\phi}{\phi^2}(\phi_{;\mu;\nu}+\frac{1}{2}g_{\mu\nu}\Box\phi)\nonumber \\-\frac{2}{\phi}(\delta\phi_{;\mu;\nu}-\delta\Gamma^\lambda_{\mu\nu}\phi_{,\lambda}+\frac{1}{2}g_{\mu\nu}\Box\delta\phi-\frac{1}{2}g_{\mu\nu}g^{\lambda\rho}\delta\Gamma^\alpha_{\lambda\rho}\phi_{,\alpha}+\frac{1}{2}\phi_{;\lambda;\rho}(g^{\lambda\rho}h_{\mu\nu}-g_{\mu\nu}h^{\lambda\rho}))
%\end{eqnarray}
%Here we have assumed that there is no fluctuations of %matter $(\delta T_{\mu\nu}=0)$. Similar calculations have %been done using an Einstein frame line element in %\cite{zimmerman}. 
The metric and the scalar field fluctuations produce the gravitational wave that is detected by detectors situated in the asymptotic region where the background spacetime can be approximated as flat. The scalar field fluctuation produces a breathing mode polarization in the GWs. The strain excitation in a single detector is a combination of the projections of the gravitational waveforms corresponding to the different polarization states incident on it~\cite{isi}. In general it is not possible to isolate with quadrupolar detectors even the two polarization components predicted in General Relativity from observations of a single detector~\cite{Pai:2000zt}, let alone the five polarization states, which is the maximum number of non-degenerate states that metric theories of gravity are allowed~\cite{will}. With at least five linearly independent detectors it is possible to resolve these five polarizations from transient signals~\cite{will,dhurandhartinto}. However, that solution is beyond the scope of this work; rather, here 
%Multiple detectors can be used to isolate each mode of polarization of the gravitational wave. We 
we shall consider only the breathing mode, and use the QNMs calculated to find the errors on the estimated metric parameters using a Fisher matrix analysis. 

We now proceed towards obtaining the  time-domain profiles and the QNMs, 
which will provide inputs for parameter
estimation using GW data.

\section{Time-domain profile and quasinormal modes}

\noindent To begin, let us first look at the time-domain profile of the 
scalar field and then find the quasi-normal modes.

\subsection{Time domain profile}
The time evolution of the scalar field is obtained 
by directly integrating the differential equation following the method described in \cite{qnmreview1},
\cite{gundlach}. We can write the scalar field equation without imposing the stationary ansatz as,\\
    \begin{align}
     \frac{\partial^{2}\psi_{lm}}{\partial t^{2}}-\frac{\partial^{2}\psi_{lm}}{\partial r_{*}^{2}}+V_l(r_{*})\psi_{lm}=0
    \end{align}
Rewriting the wave equation in terms of light cone coordinates, $du=dt-dr_*$ and $dv=dt+dr_*$ we obtain,
\begin{align}\label{waveeqnuv}
    \left( 4\frac{\partial^2}{\partial u\partial v}+V_l(u,v)\right)\psi_{lm}=0
\end{align}
In these coordinates, the time evolution operator is,\\ 
\begin{align}
    \exp\lb h\frac{\partial}{\partial t}\rb &=\exp\left(h\frac{\partial}{\partial u}+h\frac{\partial}{\partial v}\right)\\
    %&=-1+\exp(h\frac{\partial}{\partial u})+\exp(h\frac{\partial}{\partial v})+h^2\frac{\partial^2}{\partial u\partial v}+\frac{h^3}{2}(\frac{\partial^3}{\partial u^2\partial v}+\frac{\partial^3}{\partial u^2\partial v})+O(h^4)\\
    =-1+&\exp\lb h\frac{\partial}{\partial u}\rb+\exp\lb h\frac{\partial}{\partial v}\rb +\frac{h^2}{2}\left(\exp\left(h\frac{\partial}{\partial u}\right)+\exp\left(h\frac{\partial}{\partial v}\right)\right)\frac{\partial^2}{\partial u\partial v}+O(h^4)\nonumber
\end{align}
By acting this operator on $\psi_{lm}$ and using (\ref{waveeqnuv}), we arrive at
\begin{align}
    \psi_{lm}(u+h,v+h)=&\psi_{lm}(u+h,v)+\psi_{lm}(u,v+h)-\psi_{lm}(u,v)\nonumber\\ &-\frac{h^2}{8}V_l(u,v)\left(\psi_{lm}(u+h,v)+\psi_{lm}(u,v+h)\right)+O(h^4)
\end{align}
Using the above equation, we can calculate the values of $\psi_{lm}$ inside the square which is built on the two null surfaces $u=u_0$ and $v=v_0$, starting from the initial data specified on them.
% \begin{figure}[H]
 % \includegraphics[scale=1]{nullgrid}
  %\caption{The integration grid. Each cell represents an integration step. Initial data are specified on the left and bottom sides of the square.}
 %\end{figure}
The plots (Figure 2) below show the time domain profile of the field calculated for various parameter values. For the $v=0$ null line, a Gaussian profile of width $14$ centered at $u=10$ is assumed. On the $u=0$ line we have assumed  constant data. 
The field has been calculated in the region $0<u<200$ and $0<v<200$ with a 
step-size of $h=0.1$. Figure 2 shows the time domain profiles for various
values of the parameters.
\begin{figure}[h]
	\centering
	\begin{subfigure}[b]{0.35\textwidth}
		\centering
		\includegraphics[width=\textwidth]{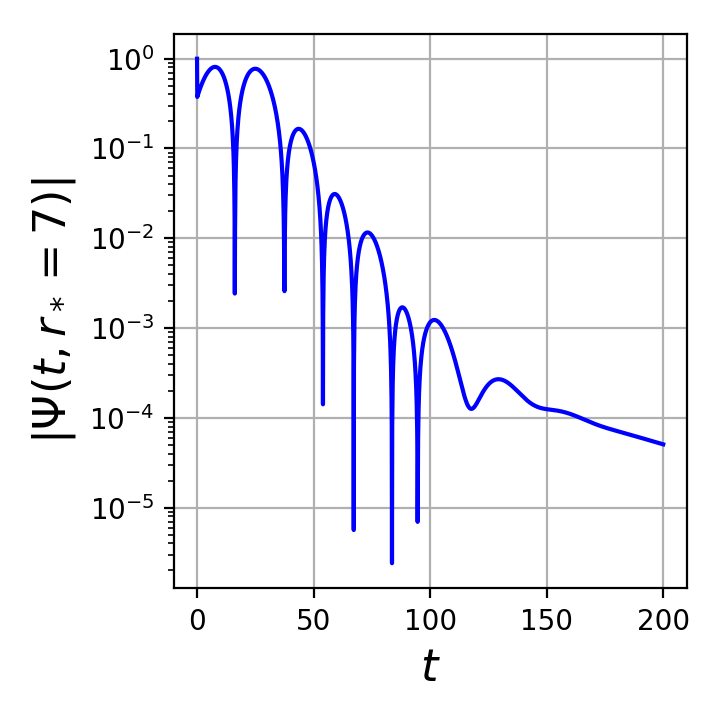}
		\caption{$l=0$}
	\end{subfigure}%\hspace{-1em}
	\begin{subfigure}[b]{0.35\textwidth}
		\centering
		\includegraphics[width=\textwidth]{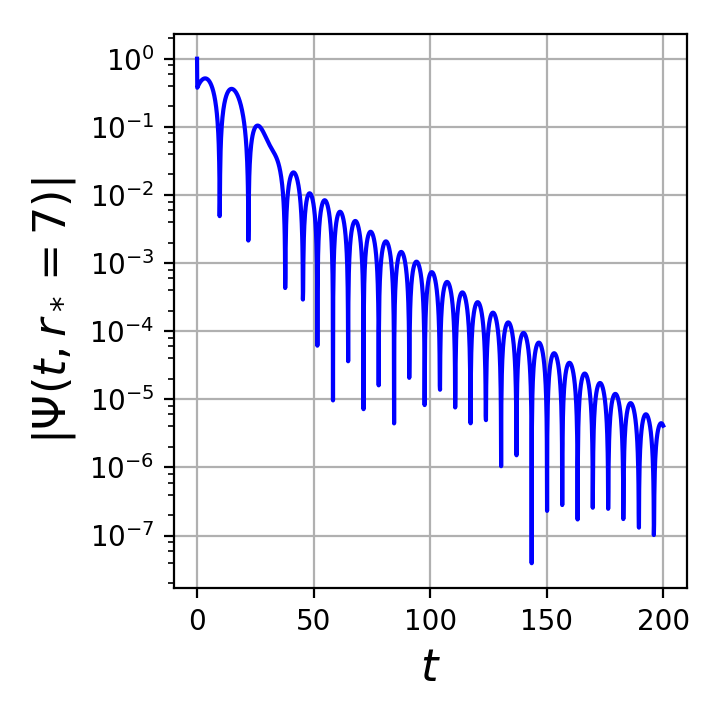}
		\caption{$l=1$}
	\end{subfigure}%\hspace{-1.5em}
	\begin{subfigure}[b]{0.35\textwidth}
		\centering
		\includegraphics[width=\textwidth]{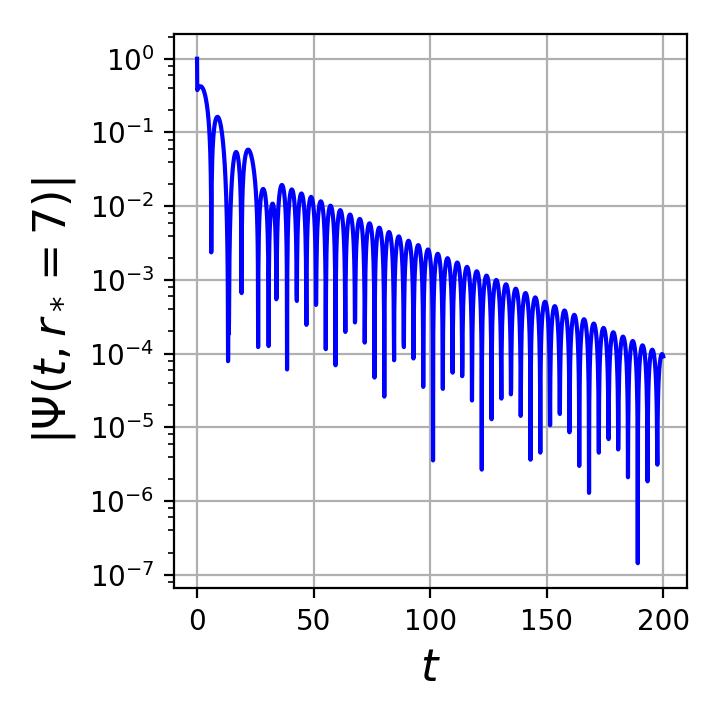}
		\caption{$l=2$}
	\end{subfigure}%
	\caption{Time domain profile and quasinormal
          ringdown. Profiles have been calculated for
          $\kappa=0.5=\lambda$, $M=1$ and $l=0,1,2$ at
          $r_*=7$. Initial conditions are
          $\psi_{lm}(u,0)=\exp\left[-\frac{(u-10)^2}{100}\right]$ and
          $\psi_{lm}(0,v)=1$. The integration grid is $u,v\in (0,200)$
          with $h=0.1$. (a) l=0, (b) $l=1$ and (c) $l=2$.}
	\label{timedomprof}
\end{figure}
\subsection{Quasi-normal modes}
The quasinormal modes, first discussed in \cite{vishu}, are defined as complex eigenfrequencies of the wave 
equation (\ref{mastereqn}) which satisfies the boundary conditions 
$\psi_{lm}\sim e^{\pm i \omega r_*}$ in the asymptotic regions $r_*\rightarrow \pm\infty$ \cite{qnmreview1,qnmreview2}. At the throat, we impose the continuity of $d\psi_{lm}/dr_*$. Since the potential is symmetric about $r_*=0$, the eigenfunctions should be either symmetric or antisymmetric. Thus, we get two families of QNMs corresponding to the initial conditions $\psi_{lm}(0)=0$ and $\psi_{lm}'(0)=0$, that can be obtained by a direct integration of the wave equation \cite{paniqnm}.
%$\left.\frac{d\Psi}{dr_*}\right|_{r_*=0}=0$.

\noindent In the asymptotic region we solve (\ref{mastereqn}) by expanding $\psi_{lm}$ as a power series upto a finite but arbitrary order,
\begin{align}\label{infseries}
    \psi_{lm} =e^{k r_*}\sum_{n=0}^{N}\frac{a_n}{r^n}
    =e^{k r}r^{kM(1+\lambda)}\sum_{n=0}^{N}\frac{\bar{a}_n}{r^n}
\end{align}
where we have used the asymptotic expansion of $r_*(r)$ in (\ref{infseries}). By substituting (\ref{infseries}) into (\ref{mastereqn}) and expanding it in terms of $1/r$ we can solve for the coefficients $\bar{a}_1,\bar{a}_2,\bar{a}_3,...$ in terms of $\bar{a}_0$. Near $2M$ we expand $\psi_{lm}$ as,
\begin{align}\label{horseries}
    \psi_{lm}=\sum_{n=0}^{N'}b_{n/2}(r-2M)^{n/2}
\end{align}
Using the same method as stated above, $b_1,b_{3/2},b_2...$ can be solved in terms of $b_0$ and $b_{1/2}$. The tortoise coordinate, given in (\ref{rstar}), near $r=2M$ can be written 
as,
\begin{align}\label{rstarseries}
    r_*=c_{1/2}(r-2M)^{1/2}+c_1(r-2M)+c_{3/2}(r-2M)^{3/2}+....
\end{align}
From (\ref{horseries}) and (\ref{rstarseries}) we get,
\begin{align}
    \left.\frac{d\psi_{lm}}{dr_*}\right|_{r_*=0}=\frac{b_{1/2}}{c_{1/2}}
\end{align}
Thus, we calculate the QNMs by integrating the wave equation from $r_0=2M(1+\delta)$ (where $\delta\||1$) to a large value of $r$ and comparing it with the two independent solutions obtained 
by substituting $k=\pm i\omega$ in (\ref{infseries}). This gives two families of QNMs by 
starting with either $b_0=0$ or $b_{1/2}=0$, which corresponds to the initial conditions $\psi_{lm}(0)=0$ and $\psi_{lm}'(0)=0$ respectively.
\begin{figure}[h]
	\centering
	\begin{subfigure}[b]{0.495\textwidth}
		\centering
		\includegraphics[width=\textwidth]{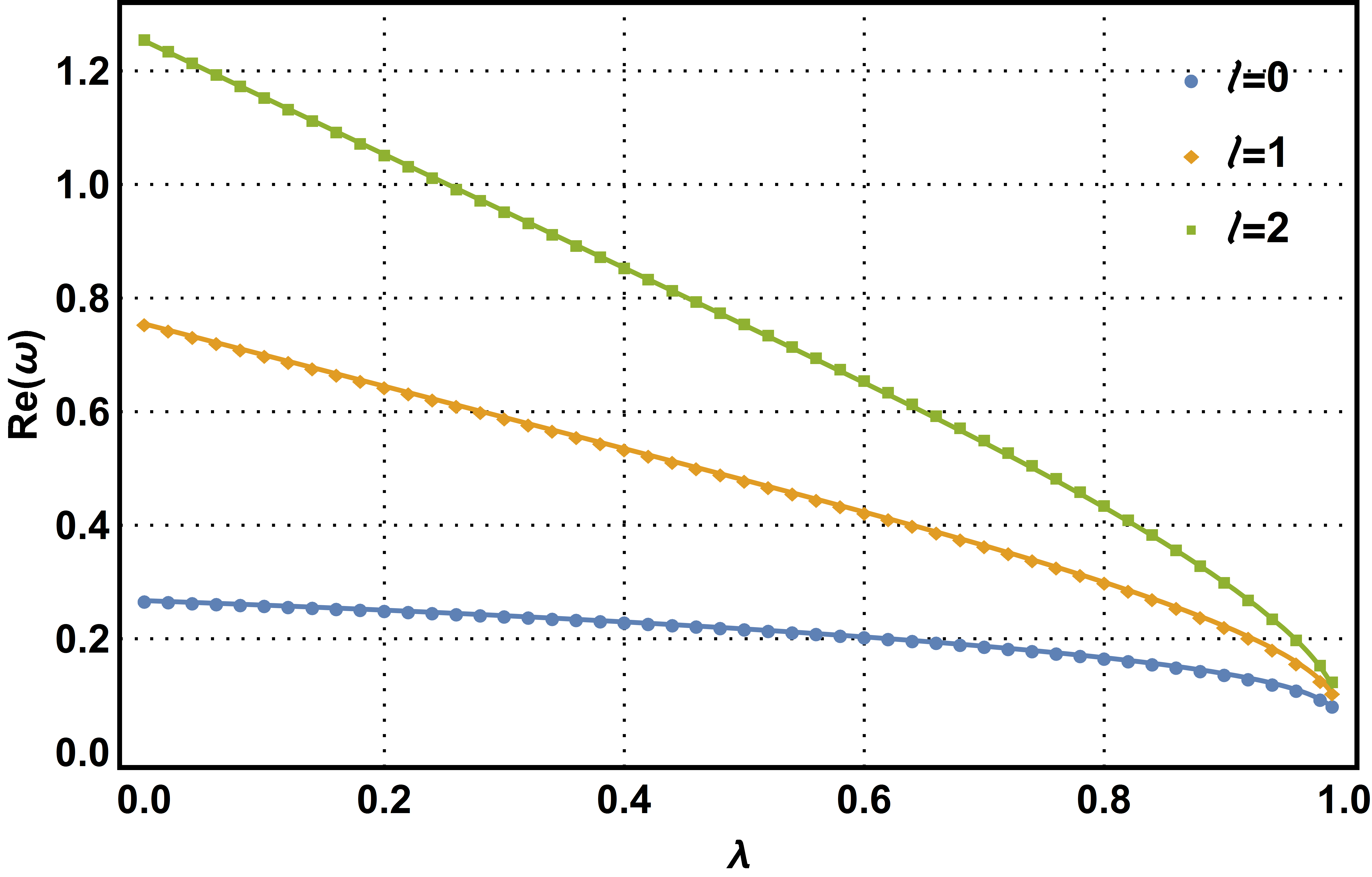}
		%\caption{$\kappa=\lambda=0.5, \mu=0$}
	\end{subfigure}\hspace{-0.3em}
	\begin{subfigure}[b]{0.495\textwidth}
		\centering
		\includegraphics[width=\textwidth]{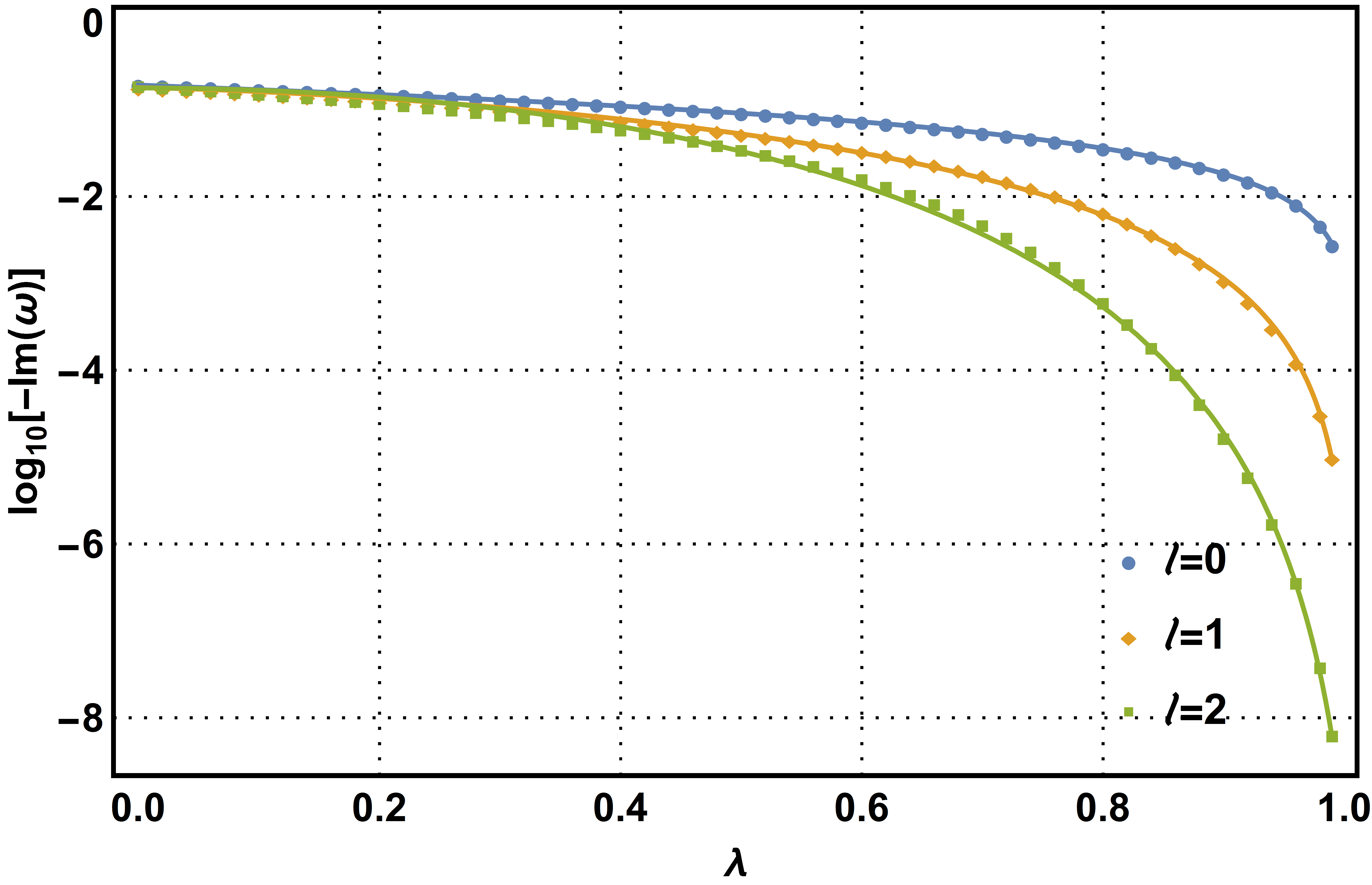}
		%\caption{$L=0, \lambda=1, \mu=0$}
	\end{subfigure}
	\caption{Real and imaginary parts of QNFs are plotted w.r.t $\lambda$. Here $M=1$ and $\kappa=1-\lambda$. Frequencies for all other values of $\kappa$, $\lambda$ and $M$ can be obtained from this data. The smooth curve joining the data points is the analytical fit which is obtained in the next section.}
	\label{qnmplot}
\end{figure}

%\noindent Figure \ref{qnmplot} show the variation of the real and imaginary parts of the $L$=0,1 and 2 mode with $\lambda$. When $\lambda=1$, the metric reduces to Schwarzschild metric. $\kappa+\lambda=1$ is assumed and $M=1$.

%\noindent The value of $\omega$ for $\lambda=0.5$, calculated by direct 
%integration method is {\bf $0.752257-0.0328503i$}. The value of $\omega$ has also been calculated by fitting the time domain profile with a damped sinusoid, using the Prony method. This gives {\bf $0.75002394-0.03201102i$}. Thus, the QNMs calculated by both methods are in good agreement.\\

\noindent The values of $\omega$ vs $\lambda$ for $l=0,1,2$ are shown in the 
Figure (\ref{qnmplot}). Here, as before, $\kappa+\lambda=1$ and $M=1$ is assumed. $\lambda=0$ is an
ultrastatic spacetime and $\lambda=1$ is the Schwarzschild limit. 
$\omega$ is given in geometric units.
Quasinormal frequencies can also be calculated directly from the time
domain profile (Figure \ref{timedomprof}) by fitting it with damped
sinusoids using Prony fit method \cite{qnmreview1} (see Fig.~\ref{pronyfit}). Few of the frequencies obtained through both direct integration and Prony fit methods are given in Table \ref{pronytable}. Both the methods are found to be consistent with each other.

\begin{table}[h]
\begin{tabular}{|c|c|c|c|c|c|}
\hline
$\kappa$ & $\lambda$ & $l$& $\omega_{prony}$ & $\omega_{DI}$\\
\hline
0.9 & 0.1 & 2 & $1.15101310-0.14169729i$ & $1.15097690 - 0.14490430i$\\ \hline
0.7 & 0.3 &2& $0.94816165-0.08240778i$ & $0.94989921 - 0.08342129i$ \\ \hline
0.5 & 0.5 &2& $0.75002394-0.03201102i$ & $0.75225700 - 0.03285035i$ \\ \hline
0.4 & 0.6 &2& $0.65230556-0.01514451i$ & $0.65232260 - 0.01512223i$ \\ \hline
0.2 & 0.8 &2& $0.43641027-0.00040061i$ & $0.43221577 - 0.00056867i$ \\ \hline
\end{tabular}
\caption{QNMs computed through Prony fit of time domain profile and direct integration}\label{pronytable}
\end{table}
\begin{figure}[h]
\centering
\includegraphics[scale=0.7]{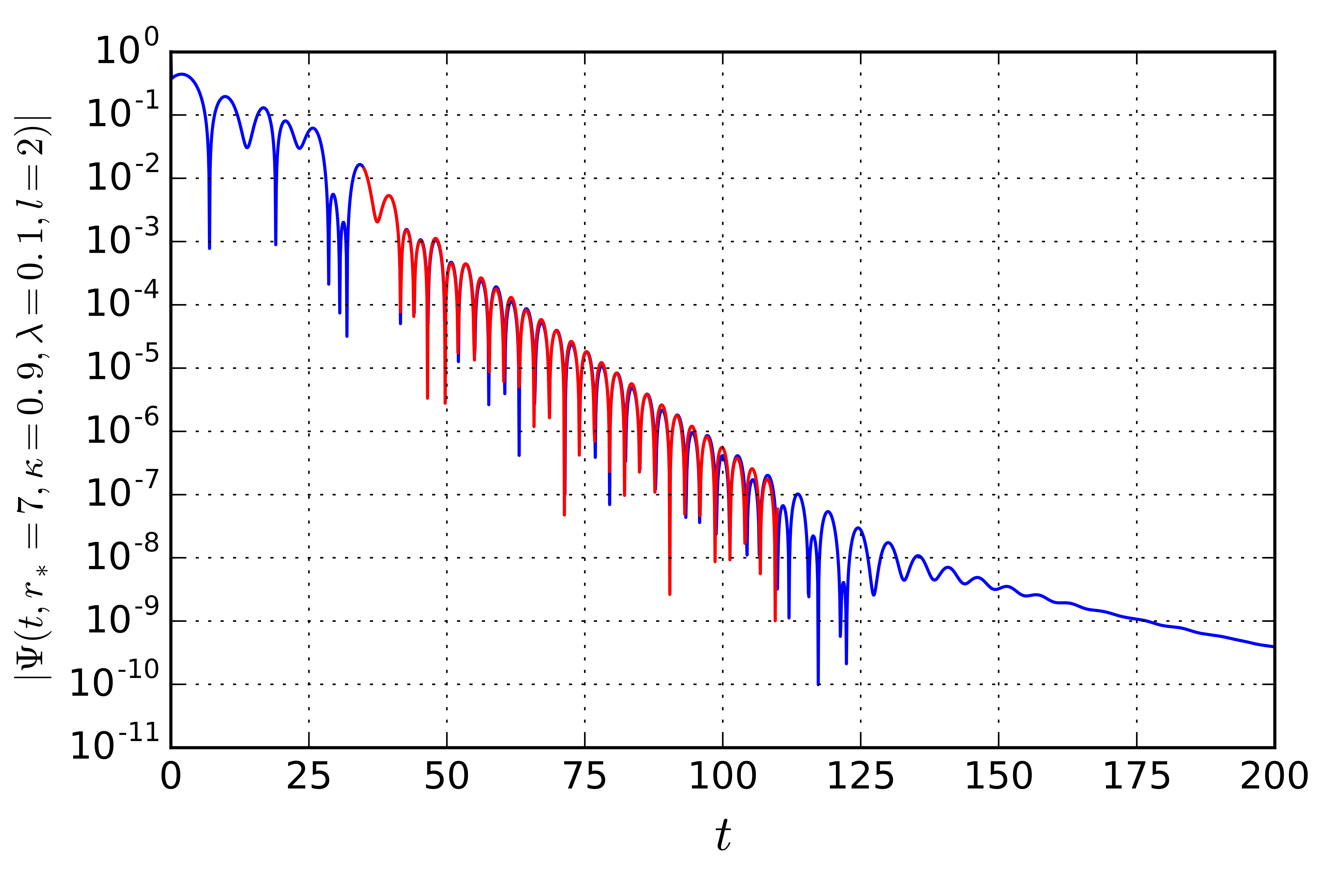}
\caption{An example of Prony fit of time domain profile. The data is fitted with four complex frequencies.}
\label{pronyfit}
\end{figure}

\subsection{Approximate Analytic Fit}
\noindent In order to perform parameter estimation on the gravitational waves comprised of QNMs, we need to calculate the derivatives of QNMs with respect to $\frac{\kappa}{\lambda}$ and $M$. For this, we construct an approximate model for $\omega$  as follows (here $\kappa=1-\lambda, M=1$),
\begin{align}
\begin{split}
\omega_{re}=(a-b\lambda)(1-\lambda^n)^m \\
\ln(-\omega_{im})=c+d\ln\left(1-p\lambda^q\right)\\
\omega=\omega_{re}+i\omega_{im}
\end{split}
\end{align}
%Since at infinity $d\tau=(\kappa+\lambda)dt$, the frequency measured will be $\omega'=\frac{\omega}{\kappa+\lambda}$.
The values of $a,b,c,d,m,n,p,q$ were calculated using {\sf NonlinearModelFit} 
in {\em Mathematica} \cite{Mathematica}. For $l=0$,
\begin{equation}
\begin{split}
     \omega_{re}&=(0.27-0.08\lambda)(1-\lambda^{2.36})^{0.24}\\
     \omega_{im}&=-0.19(1-0.99\lambda^{0.94})^{1.02}
\end{split}
\end{equation}
For $l=1$,
\begin{equation}
\begin{split}
     \omega_{re}&=(0.75-0.55\lambda)(1-\lambda^{7.95})^{0.28}\\
     \omega_{im}&=-0.18(1-0.99\lambda^{1.41})^{2.63}
\end{split}
\end{equation}
and for $l=2$, we have
\begin{equation}
\begin{split}
     \omega_{re}&=(1.25-\lambda)(1-\lambda^{8.87})^{0.33}\\
      \omega_{im}&=-0.18(1-0.96\lambda^{1.99})^{6.05}
\end{split}
\end{equation}
Figure \ref{qnmplot} shows the plots of analytical model of the QNMs. Since at infinity $d\tau=(\kappa+\lambda)dt$, the frequency of the signal measured by an observer at the asymptotic region will be $\nu=\frac{\omega_{re}}{2\pi(\kappa+\lambda)}$ and the time constant will be $\tau=\frac{\kappa+\lambda}{|\omega_{im}|}$.
Thus, for $l=0$,
\begin{equation}
\label{eq:L0}
\begin{split}
    \nu=&\frac{8628.13}{M}\left(1-0.29\frac{\lambda}{\lambda+\kappa}\right)\left(1-\left(\frac{\lambda}{\lambda+\kappa}\right)^{2.36}\right)^{0.24}Hz\\
    \tau=&\frac{M}{38908.58}\left(1-0.99\left(\frac{\lambda}{\lambda+\kappa}\right)^{0.94}\right)^{-1.02} s
\end{split}
\end{equation}
For $l=1$,
\begin{equation}
\label{eq:L1}
\begin{split}
    \nu=&\frac{24375.84}{M}\left(1-0.73\frac{\lambda}{\lambda+\kappa}\right)\left(1-\left(\frac{\lambda}{\lambda+\kappa}\right)^{7.95}\right)^{0.28}Hz\\
    \tau=&\frac{M}{36290.32}\left(1-0.99\left(\frac{\lambda}{\lambda+\kappa}\right)^{1.41}\right)^{-2.63} s
\end{split}
\end{equation}
For $l=2$,
\begin{equation}
\label{eq:L2}
\begin{split}
    \nu=&\frac{40478.89}{M}\left(1-0.80\frac{\lambda}{\lambda+\kappa}\right)\left(1-\left(\frac{\lambda}{\lambda+\kappa}\right)^{8.87}\right)^{0.33}Hz\\
    \tau=&\frac{M}{36026.46}\left(1-0.96\left(\frac{\lambda}{\lambda+\kappa}\right)^{1.99}\right)^{-6.05} s
\end{split}
\end{equation}
$M$ is in units of solar mass, $\nu$ is in Hz and $\tau$ is in seconds. Moreover, the validity of the above fits is verified for $\frac{\kappa}{\lambda}$ larger than a few times 0.001.
%$\in\left(0.01,\infty\right)$. 

The gravitational-wave strain in a detector is a linear function of the various polarization components the theory may allow,
\begin{equation}
h(t) = \sum_A F^A h_A\,,
\end{equation}
where $A$ is the polarization index, $h_A$ are GW polarization
components, and the coefficients $F^A$ are the
antenna-pattern functions that are determined by how well the
polarization components project on the GW detector. The $F^A$ depend
on the sky-position angles $(\vartheta,\varphi)$ of the source and the polarization angle 
%$\psi$ 
of the gravitational wave, in general. In our case, the source is the wormhole studied here. The contribution to the detector strain from the breathing mode alone of such a source will be considered in this work, and is given by
\begin{equation}
\label{eq:dampedsinusig}
h(t)=A\sin(2\pi\nu t)e^{-t/\tau}\,,
\end{equation}
where the strain amplitude $A$ contains the breathing-mode antenna pattern \cite{isi}
\begin{equation}
F^b = -\frac{1}{2}\sin^2\vartheta \cos 2\varphi\,.
\end{equation}
Above, $\vartheta$ and $\varphi$ are the 
%right ascension and declination, respectively, of the 
polar angle and azimuthal angle, respectively, that define the
sky-position of the source in a coordinate system where the two arms
of the quadrupolar detector are the $x$ and $y$ axes. Therefore, the
strength of the detector signal, which depends on $h$ linearly, will
vary across the sky even if the rest of the wormhole parameters remain
unchanged. Below, for estimating parameter errors, we will take the
source to be located along the $x$ or $y$ arm of the detector, i.e., $\vartheta = \frac{\pi}{2}$ and $\varphi = 0$ or $\pi/2$.

%In the absence of detector noise, 
If a loud enough damped-sinusoid strain signal~\eqref{eq:dampedsinusig}  is observed in a detector, the parameters of the wormhole can be deduced from a straightforward Fourier transform. For example, by an observation of the $l=0$ signal in Eq.~\eqref{eq:L0}, one can infer from its measured central frequency $\nu$ and the time-constant $\tau$, the mass $M$ and geometry parameter $\kappa/\lambda$. When the signal is strong enough to allow the observation of multiple modes -- the higher modes will get progressively weaker inherently, but their signal-to-noise ratio will also depend on the amplitude of the detector noise at the mode frequency -- the multiple measured mode frequencies and time-constants can be used to perform self-consistency checks or even rule out a wormhole as the source of the signals.

Note that other sources of damped-sinusoid signals can exist in the GW
detectors, both astrophysical and terrestrial in
origin~\cite{BDGL,Bose:2016sqv}. To improve the odds of the former, it
is important to observe the commensurate signals in multiple GW
detectors~\cite{Bose:1999pj,Pai:2000zt}. But to distinguish one
astrophysical source from another, e.g., QNMs of black holes in
General Relativity~\cite{Talukder:2013ioa} or other braneworld models~\cite{Seahra:2004fg,Chakraborty:2017qve}, further comparative studies of their signals are required.
%, e.g., of the multipolar structures of signals from various astrophysical signals.

A remaining practical issue is that signals will typically be immersed in detector noise, and the measurement of any of their parameters will have errors. This is what we study next.
%, noise which we deal below

%Figure 5 show the analytic fits for the various cases mentioned above.
%\begin{figure}[h]
%	\centering
%	\begin{subfigure}[b]{0.5\textwidth}
%		\centering
%		\includegraphics[width=0.98\textwidth]{rewfit}
%		%\caption{$\kappa=\lambda=0.5, \mu=0$}
%		\label{fig:sub1}
%	\end{subfigure}\hspace{-0.5em}
%	\begin{subfigure}[b]{0.5\textwidth}
%		\centering
%		\includegraphics[width=0.98\textwidth]{imwfit}
%		%\caption{$L=0, \lambda=1, \mu=0$}
%		\label{fig:sub2}
%	\end{subfigure}
%	\caption{Approximate fit of $\omega$ vs $\lambda$ for %L=0,1,2. Dots represent the original data and the continuous lines are the approximate fitted curves}
%	\label{fig:test}
%\end{figure}

\section{Gravitational wave observations of the modes}

\noindent We use the Fisher information-matrix
formalism~\cite{Helstrom} to estimate how accurately the wormhole parameters will be measurable 
using interferometric detectors like aLIGO.
To estimate the error in $\kappa/\lambda$, we compute that
matrix for the damped-sinusoid signal (\ref{eq:dampedsinusig}) in a single aLIGO detector
at design sensitivity~\cite{aLIGOZDHP} for that parameter alone. The matrix is determined by the derivative of the signal $h$ with respect to $\kappa/\lambda$, which influences both the frequency and the damping time-constant of the signal. For this first study, we take $M$ to be known. For wormholes that result from the merger of two black holes this parameter can be estimated from the inspiral part of the signal. 
Even so, such an estimation also requires knowledge of the strength of the signal. Currently, it is not understood how large the QNM amplitude
of these wormholes can be, whether they form in binary black hole merger processes or otherwise. Therefore, for
reference we take the maximum QNM strain amplitude to be $10^{-21}$,
which is approximately the maximum amplitude of the GW150914 
signal~\cite{gwobs1}. We recognize that this choice is arbitrary. If
at a later date realistic amplitudes are deduced theoretically or
numerically, then the errors obtained in this paper should be scaled
appropriately by using those values. Finally, we invert the
information matrix to derive the estimated variance in the measured 
values of $\kappa/\lambda$~\cite{Helstrom}. 
Its square-root gives the lower bound on the statistical error in
$\kappa/\lambda$. To deduce the error for multiple statistically
independent observations, one simply replaces the information matrix for
a single observation in the above procedure with the sum of the
information matrices for multiple observations.
%~\cite{Cramer46,Rao45}. gwobs1

\begin{figure}[h]
%\label{fig:kappaVsLambdaGW150915aLIGO}
\includegraphics[scale=0.45]{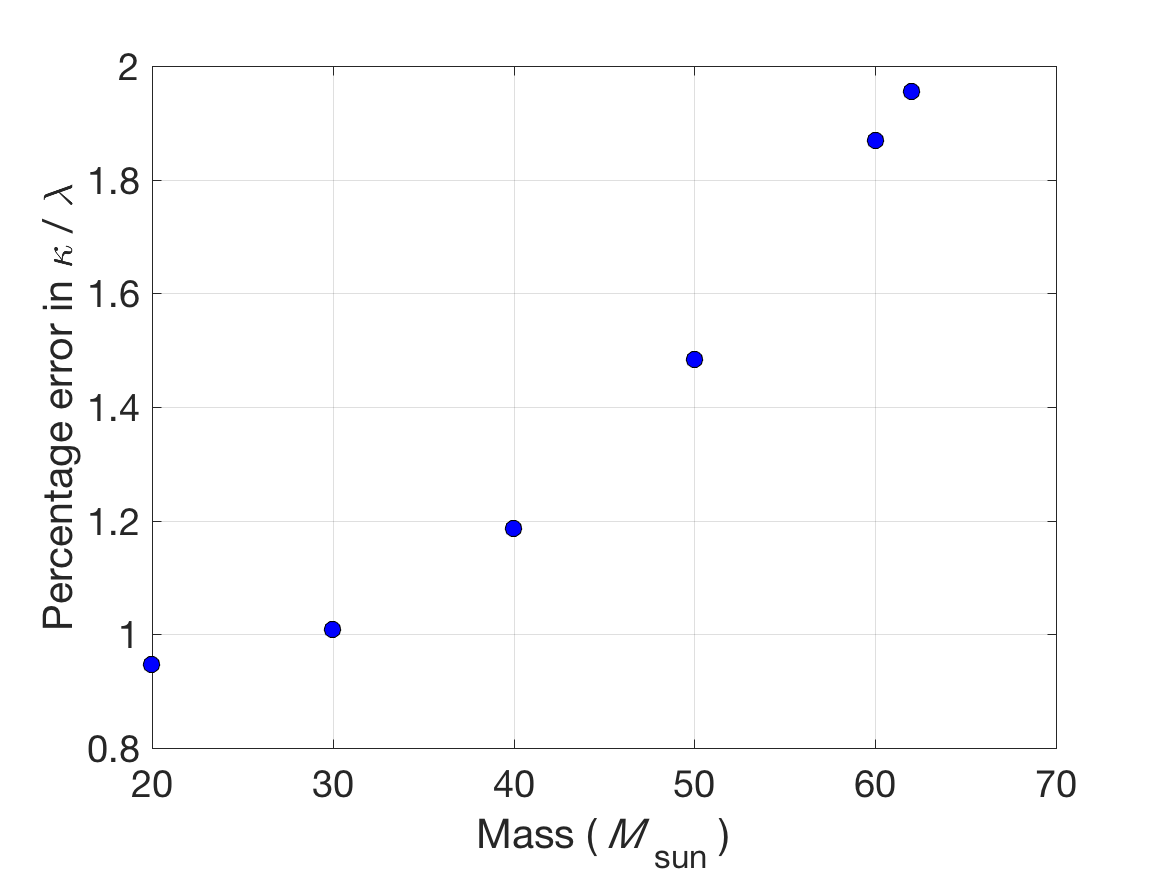}
\includegraphics[scale=0.45]{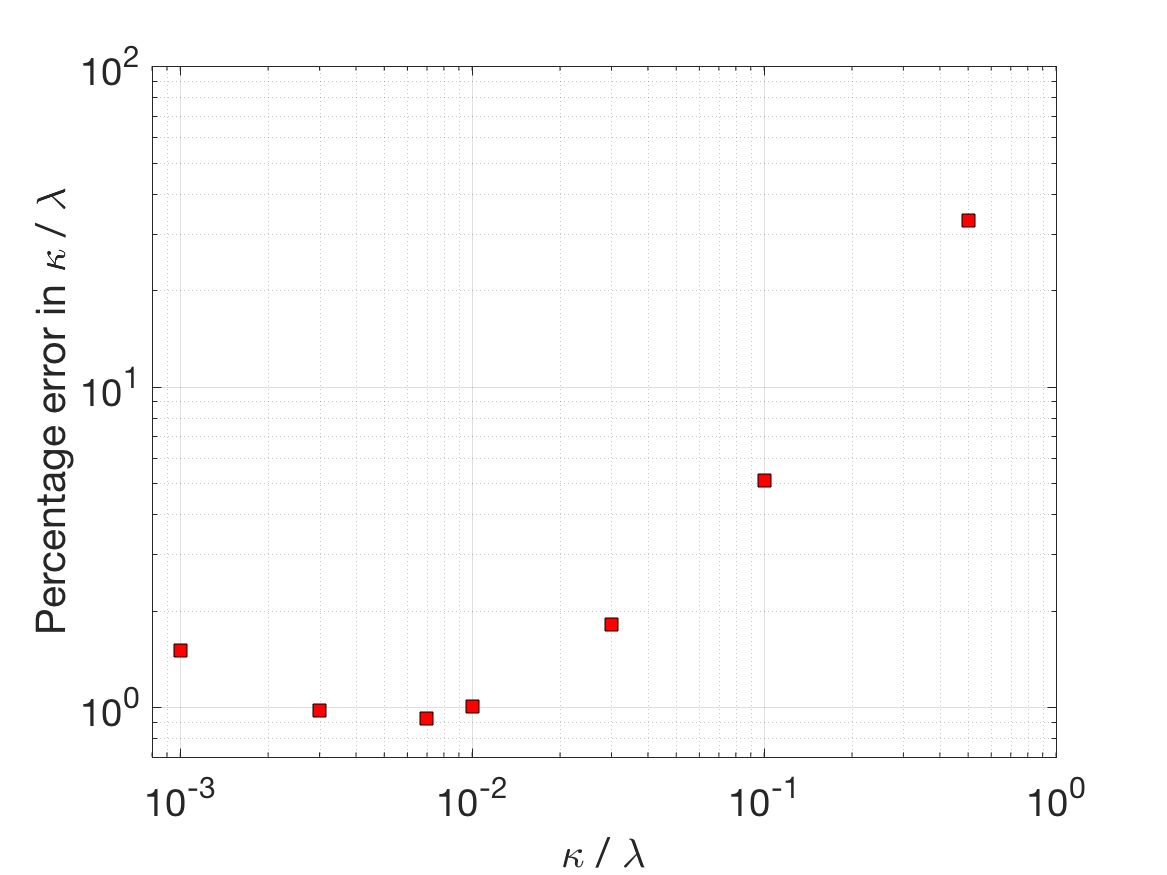}
\caption{The estimated statistical error in $\kappa / \lambda$ (in
  percentage) is shown for various values of the wormhole mass $M$ (in
  the left figure, where $\kappa / \lambda$ is kept fixed at 0.01) and
  the wormhole parameter $\kappa / \lambda$ (in the right figure,
  where $M$ is kept fixed at $30M_\odot$). These estimates were
  obtained for gravitational-wave observations of the breathing mode
  of the wormhole solution studied here with a detector like
  aLIGO. The maximum amplitude of the mode in all cases is set to be
  $10^{-21}$, which is approximately the peak amplitude of GW150914, whose final
detector-frame mass had a median value of about $68M_\odot$. The wormhole QNM frequency for $l=0$ is 64~Hz for $M=68M_\odot$ and $\kappa / \lambda = 0.1$. For comparison, the frequency of the $l=2$, $m=2$ QNM of GW150914 was at or above $\simeq 243$~Hz \cite{gwobs1}. The error in $\kappa / \lambda$ increases with $M$ (left figure) primarily because the mode frequency decreases, thereby, placing the signal in less sensitive part of the detector band. The right figure shows that the  error in $\kappa / \lambda$ initially reduces when the value of that parameter is increased. This happens because the mode frequency shifts to more sensitive parts of the detector band (from 64~Hz at $\kappa / \lambda = 3\times 10^{-3}$ to 85~Hz at $\kappa / \lambda = 10^{-2}$). For higher values of $\kappa / \lambda$, the error increases owing to decreasing time constant (and, therefore, the effective integration duration) of the mode.
}
\label{fig:kappaVsLambdaGW150915aLIGO}
\end{figure}
\begin{figure}[h]
\includegraphics[scale=0.3]{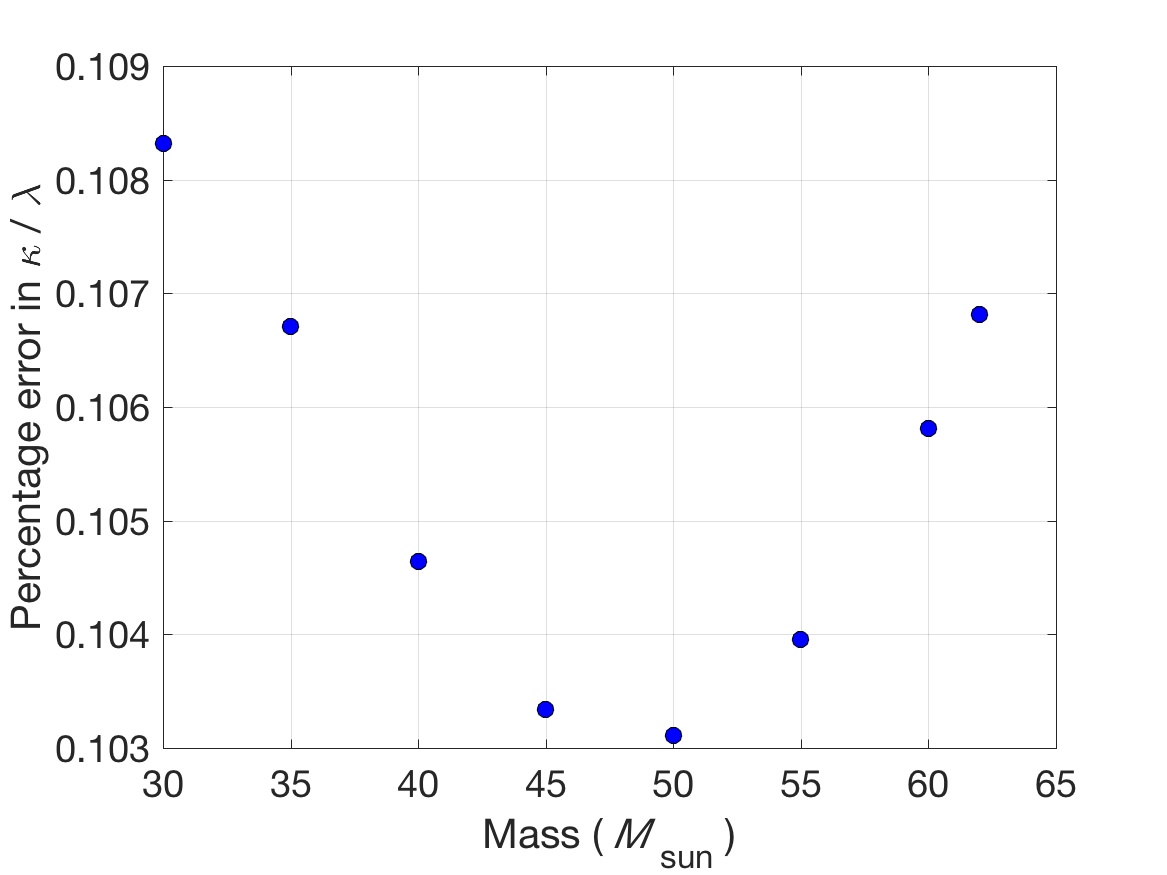}\hspace{-0.2in}
\includegraphics[scale=0.3]{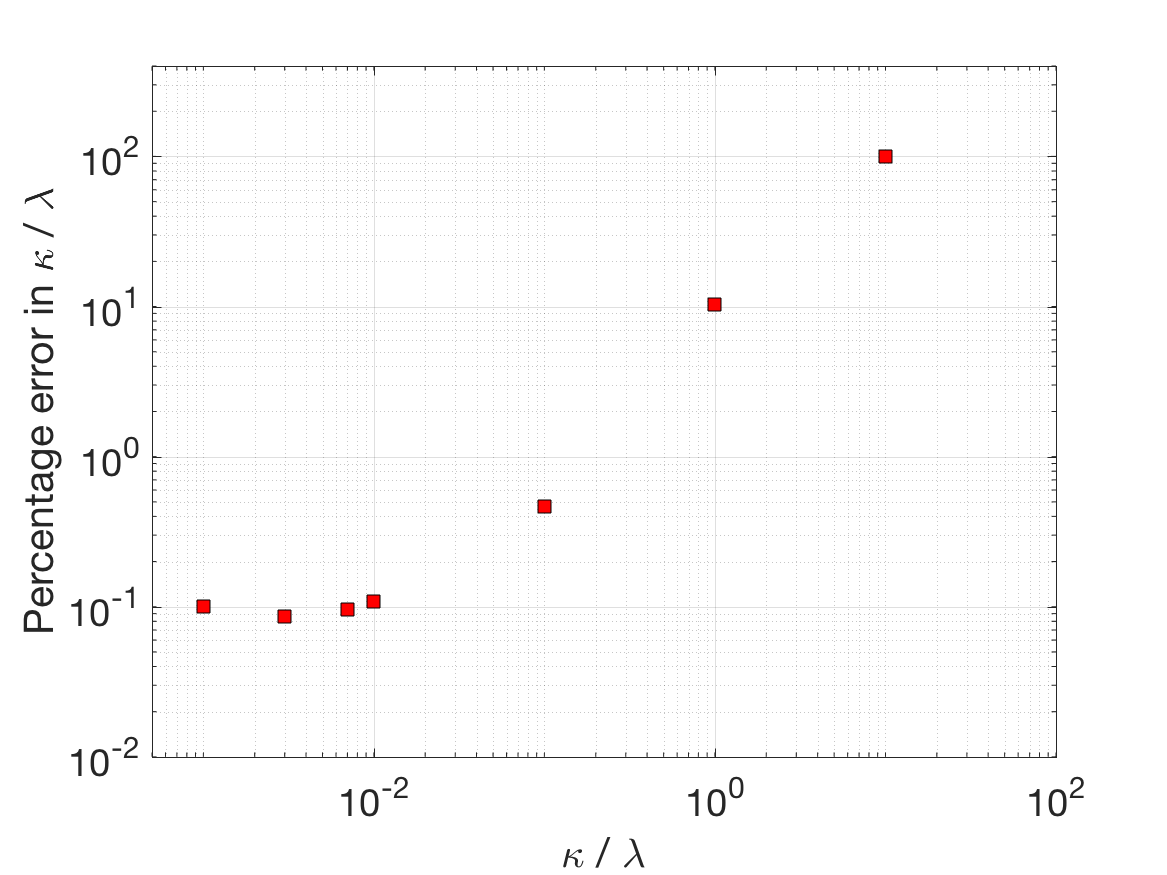}\hspace{-0.2in}
\includegraphics[scale=0.3]{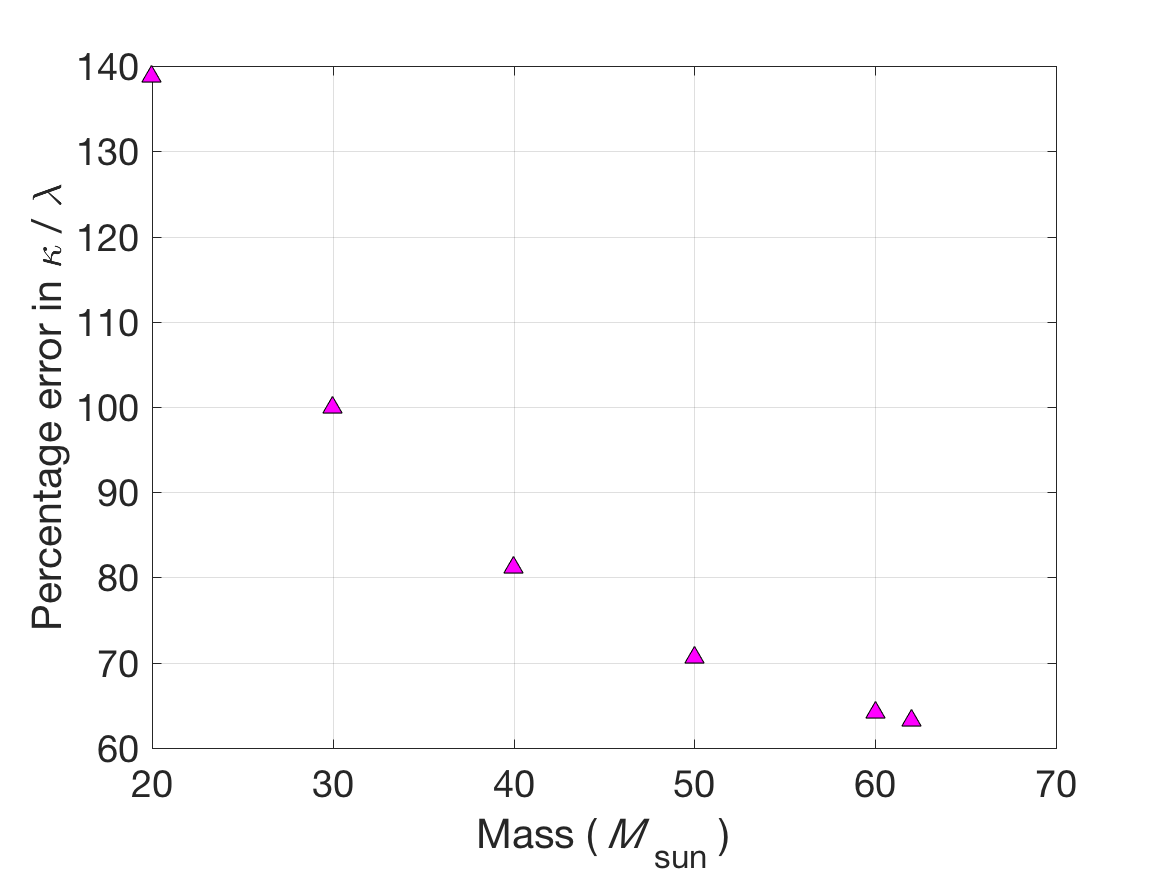}
\caption{The two left-most plots are similar to the ones shown in Fig. \ref{fig:kappaVsLambdaGW150915aLIGO}, but for the Einstein Telescope \cite{Hild}. The right-most plot 
is for $\kappa/\lambda = 10$ and the Einstein Telescope; it is interesting how the error decreases with increasing mass for the mass range shown here, before it rises 
again (not shown) owing to the signal frequency moving into the low-frequency seismic band).  Moreover, whereas $\kappa/\lambda < 1$ in Fig. \ref{fig:kappaVsLambdaGW150915aLIGO} here 
we raise that limit in the middle and right plots.}
\label{fig:ET}
\end{figure}

The parameter errors in $\kappa/\lambda$ are 
%computed for a single aLIGO detector with zero-detuned high-power noise PSD~\cite{BDGL}, and 
plotted in Figs.~\ref{fig:kappaVsLambdaGW150915aLIGO} and ~\ref{fig:ET} for observations
of the breathing mode of the wormhole described above in an aLIGO and Einstein Telescope detector, respectively.
%formed from the merger of a GW150914 like event.
%for aLIGO design sensitivity, 
This is an optimistic estimate since errors in other source
parameters, such as the signal's time of arrival and $M$, and their
covariances, which were neglected here,
%e.g., the mass (whose measurement error we took to be negligible
%here), will 
can worsen the estimation of $\kappa/\lambda$. Moreover, while the SNR
of the complete signal of GW150914 was moderately high, that of the
post-merger signal was not. Hence, the Fisher estimates, which we base
on the peak amplitude of the merger signal, must be followed up with
more reliable parameter estimation analyses; this preliminary
study therefore makes the case for adapting a more realistic approach
in a future work for estimation of $\kappa/\lambda$. Such an approach
can be the use of Monte Carlo methods, as demonstrated for binary
black hole parameter estimation in Ref.~\cite{Ajith:2009fz}, or
Bayesian methods, such as that used in Ref.~\cite{DelPozzo:2013ala}
for combining the posteriors of the tidal deformability parameter,
which describes neutron star composition, from multiple binary neutron
star coalescences. Indeed, an exercise that can be performed
is the improvement in the estimation for $\kappa/\lambda$ by combining
multiple observations. One way to do that would be to compute a joint
posterior. Another, less optimal method, would be to stack-up the power
from the multiple signals, as was done for estimating the post-merger
oscillation parameters of hypermassive neutron stars
in Ref.~\cite{Bose:2017jvk}. Since Figs.~\ref{fig:kappaVsLambdaGW150915aLIGO} and
~\ref{fig:ET} suggest a wide enough range of $\kappa/\lambda$ where its value may be measurable fairly accurately, e.g., to distinguish the wormhole geometry from Schwarzschild, it appears to be worthwhile to pursue these more sophisticated, and computationally expensive methods, in the future.

\section{Conclusions}
\noindent In summary, we obtained the following results in this work.

\noindent Assuming a two-parameter family of wormholes that arise in
a scalar-tensor theory of gravity, we have first derived their
scalar quasinormal modes using standard numerical methods. 
We have cross-checked our numerical methods with known results on QNMs in other
wormhole geometries available
in the literature~\cite{taylor}. The QNMs obtained and their
variation w.r.t. the parameters were then fitted using
methods of non-linear least square fitting. These results
were then used to estimate the accuracy with which the wormhole parameter $\kappa / \lambda$, which appears in the line element, can be measured using inputs from GW observations. For this first study,
%Our final result would hopefully be of use, if there is any attempt at  detecting a wormhole as a result of an astrophysical merger process. 
we kept things simple by considering the measurement of just one parameter
(i.e., $\frac{\kappa}{\lambda}$), while treating $M$, which is related to the throat radius, as known precisely.
%as a parameter that can be constrained independently. 
While it is true that energy conservation will require $M$ to be bounded from above by the total mass of the binary black hole merger that produces it, and that  this mass is measurable from observations of the inspiral phase, it is not clear yet how it determines $M$. This matter is left for future exploration.
%the inspiral phase and  may also be considered as an independent parameter for estimation.

\noindent Under the aforementioned assumptions, we find that for a certain range of the wormhole parameter
it would be possible to estimate its value from adequately loud
signals, if not in aLIGO, then in the Einstein Telescope (ET). For example, if the maximum amplitude of the breathing mode is set to be $10^{-21}$, which is approximately the peak amplitude of GW150914, the error in that parameter can be measured to within tens of percent for $\log_{10}(\kappa/ \lambda) \in (-3,-1)$ in an aLIGO-like detector at design sensitivity. This can be seen  in the right figure in Fig.~\ref{fig:kappaVsLambdaGW150915aLIGO}. There we set the $M= 30M_\odot$, which describes a wormhole that may form from the merger of stellar mass black holes that are not too heavy. For larger $M$, the error in $\kappa/ \lambda$ will be larger.  Moreover, for possible wormholes resulting from the merger of black holes of the type observed by LIGO and Virgo so far, one can determine $\kappa/ \lambda$ in ET to within a few to several tens of percent as well for $\kappa/ \lambda \leq 10$ (see Fig.~\ref{fig:ET}) and, therefore, distinguish them from Schwarzschild
(albeit, for non-spinning geometries). The important caveat is that these error estimates are expected to worsen when one expands the parameter space by including spin and accounts for the error in the wormhole mass and any covariances that may arise among those two parameters and $\kappa/ \lambda$. 
%From the perspective of signal-processing, the reason it becomes viable to do attain better estimates at higher values $\kappa/ \lambda$ and, correspondingly, at higher values of mode frequencies, is because the signal at those frequencies remains longer in band, thereby, allowing a better resolution of the mode-frequency's value itself. This, in turn, directly determines how accurately $\kappa/ \lambda$ can be measured.

\noindent Significantly, since mergers can leave behind remnants with non-vanishing angular momentum, it is important to extend the
results here by introducing rotation in the wormhole line element, thereby making it more realistic. 
Rotating wormholes have been studied in the literature \cite{teo}. QNMs for rotating
Ellis wormholes have been discussed in \cite{ringdown3}. For the line element used in this article, 
one would first have to generalize it
by including rotation. More importantly, one would first need to do this in the scalar-tensor theory and, subsequently, study the consequences for the WEC, if any.
This may be followed up by finding the QNMs and, thereafter, the estimates of parameter errors in possible GW observations, now using additionally the spin parameter.

\noindent The real question however is whether one can obtain a wormhole metric as
a result of an astrophysical merger process. There are some simplistic 
models of mergers which are analytic in nature \cite{emparan}. These could be
viable starting points for understanding whether a wormhole could  
be created at all in a merger.

 \section*{ACKNOWLEDGEMENTS}
 
 \noindent SA  thanks the Inter-University Centre for Astronomy and
 Astrophysics (IUCAA), Pune, India for supporting his academic visits
 to IUCAA, during 2017-18. The authors also thank Nathan
 Johnson-McDaniel for carefully reading the manuscript and 
making several useful comments. This work is supported in 
part by the Navajbai Ratan Tata Trust.

\end{document}